# HEAVY-ION BEAM ACCELERATION OF TWO-CHARGE STATES FROM AN ECR ION SOURCE

P.N. Ostroumov, K.W. Shepard, Physics Division, ANL, 9700 S. Cass Av., Argonne, IL, 60439
V.N. Aseev, Institute for Nuclear Research, Moscow 117312, Russia
A.A. Kolomiets, Institute of Theoretical and Experimental Physics, Moscow 117259, Russia

## *Abstract*

This paper describes a design for the front end of a superconducting (SC) ion linac which can accept and simultaneously accelerate two charge states of uranium from an ECR ion source. This mode of operation increases the beam current available for the heaviest ions by a factor of two. We discuss the 12 MeV/u prestripper section of the Rare Isotope Accelerator (RIA) driver linac including the LEBT, RFQ, MEBT and SC sections, with a total voltage of 112 MV.

The LEBT consists of two bunchers and electrostatic quadrupoles. The fundamental frequency of both bunchers is half of the RFQ frequency. The first buncher is a multiharmonic buncher, designed to accept more than 80% of each charge state and to form bunches of extremely low longitudinal emittance (rms emittance is lower than 0.2 $\pi$·keV/u·nsec) at the output of the RFQ. The second buncher is located directly in front of the RFQ and matches the velocity of each charge-state bunch to the design input velocity of the RFQ. We present full 3D simulations of a two-charge-state uranium beam including space charge forces in the LEBT and RFQ, realistic distributions of all electric and magnetic fields along the whole prestripper linac, and the effects of errors, evaluated for several design options for the prestripper linac. The results indicate that it is possible to accelerate two charge states while keeping emittance growth within tolerable limits.

## 1 INTRODUCTION

The Rare Isotope Accelerator (RIA) Facility requires a 1.3 GeV linac which would accelerate the full mass range of ions and would deliver ~400 kW of uranium beam at an energy of 400 MeV per nucleon [1,2]. The driver would consist of an ECR ion source and a short, normally-conducting RFQ injector section which would feed beams of virtually any ion into the major portion of the accelerator: an array of more than 400 superconducting (SC) cavities of seven different types, ranging in frequency from 57.5 to 805 MHz. The linac contains two stripping targets, at 12 MeV/u and 85 MeV/u, for the uranium beam. A novel feature of the linac is the acceleration of beams containing more than one charge state [3,4]. The front end of the RIA driver linac consists of a ECR ion source, a LEBT, a 57.5 MHz RFQ, a MEBT and a section of SRF drift-tube linac.

The present-day performance of ECR ion sources, and considerations based on fundamental limiting processes in the formation of high-charge state uranium ions in such sources, indicate that uranium beam intensities as high as 7 p$\mu$A in a single charge state of 29+ or 30+ are unlikely to be obtained in the near future. Such a high current is required in order to produce the RIA driver linac design goal of a 400 kW uranium beam, even if we assume multiple charge state beam acceleration following the first stripper.

This paper discusses in detail a solution to this limitation. It doubles the heavy-ion beam intensity by accepting two charge states from the ion source.

## 2 DESIGN OF THE FRONT END

### *2.1 LEBT*

The LEBT is designed for the selection and separation of the required ion species and the acceptance of single- or two-charge states by the following RFQ structure. The first portion of the LEBT is an achromatic bending magnet section for charge to mass analysis and selection. For the heaviest ions, such as uranium ions, the transport system must deliver to the entrance of the first buncher a two-charge-state beam with similar Twiss parameters for both charge states. The design features of the two-charge selector will be discussed elsewhere. The ECR is placed on a high voltage platform. The voltage $V_0$=100 kV is adequate to avoid space charge effects in the LEBT and RFQ and to keep the RFQ length to less than 4 m.

A simplified layout of the second part of the LEBT is shown in Fig. 1. This part of the LEBT solves the following tasks: a) Beam bunching by a four-harmonic external buncher $B_1$ (the fundamental frequency is 28.75 MHz); b) Velocity equalization of two different charge states by the buncher $B_2$, operating at 28.75 MHz; c) Charge-insensitive transverse focusing of the 2-charge state beam and matching to the RFQ acceptance by the electrostatic quadrupoles $Q_1$-$Q_8$.

The reference charge state for the design of the LEBT, RFQ and MEBT is 29.5. The RFQ injector is designed to accelerate any beam from protons to uranium to a velocity v/c = 0.01893 at the exit of the RFQ.

The computer code COSY [5] was used in order to design and optimize the LEBT by taking into account terms through third order. The final beam dynamics simulation has been performed by the DYNAMION code [6] where the equations of motion are solved in a general approximation using realistic 3D electrostatic fields of the quadrupoles and rf bunchers, including space charge forces for the multi-charged ion beams. Realistic 3D fields for the electrostatic quadrupoles were calculated by the SIMION code [7]. The 3D field distributions have been used with both the COSY and DYNAMION codes. Figure 2 shows beam envelopes along the LEBT optimized by COSY. The total normalized emittance at the exit assumes an ion source emittance of 0.5 π·mm·mrad. After careful optimization, including third order terms, the rms emittance growth is less than 7% in the horizontal plane. In the vertical plane there is no observable rms emittance growth. Figure 3 shows the transformation of the beam image in longitudinal phase space. The first multi-harmonic buncher modulates the beam velocity of the two charge states as shown in Fig. 3a. The drift space between the two bunchers (see Fig. 3) is chosen from the expression

$$L_{12} = \lambda\sqrt{\frac{2eV_0}{Am_u}} \cdot \frac{\sqrt{q_0(q_0-1)}}{\sqrt{q_0}-\sqrt{q_0-1}},$$

where $\lambda$ is the wavelength of the RFQ frequency, $m_u$ is the atomic unit mass. This drift space separates in time the bunches of different charge states $q_0$ and $q_0$-1 by 360° at the RFQ frequency. The voltages of the multi-harmonic buncher have been optimized, together with the RFQ parameters, in order to obtain a total efficiency above 80%, while minimizing the longitudinal emittance for each charge state. The second buncher is used to equalize the velocities of the 2 charge states (see Fig. 3c).

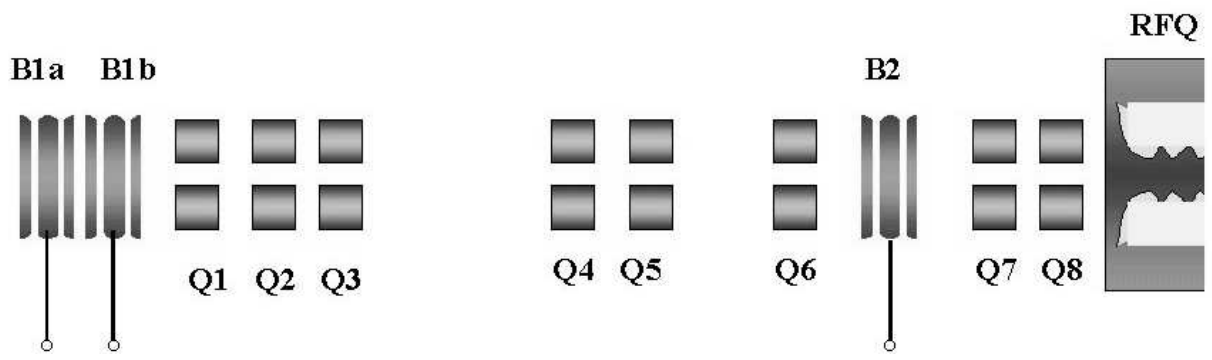

Figure 1: Layout of the LEBT.

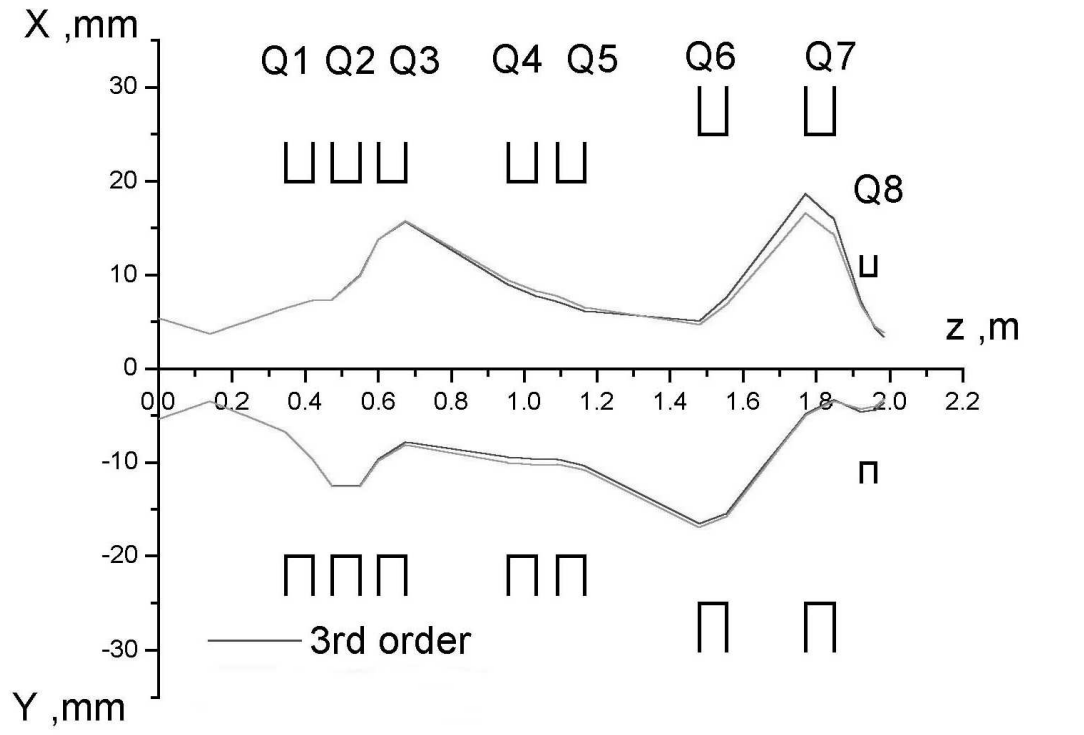

Figure 2: Beam envelopes in the LEBT.

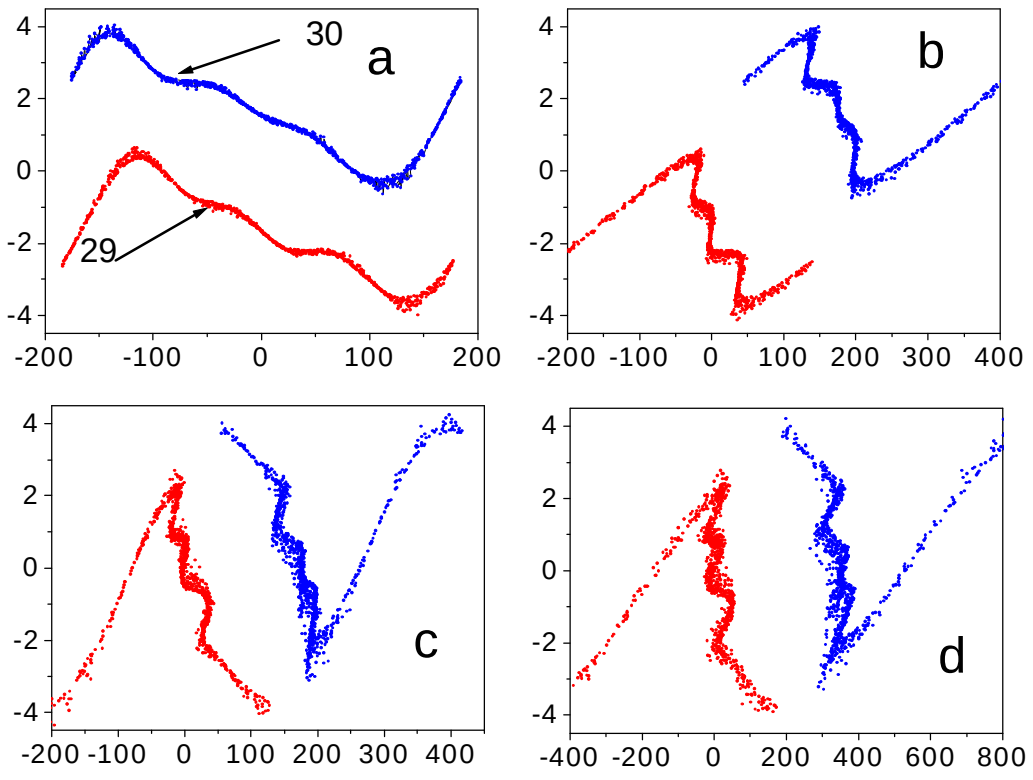

Figure 3: Longitudinal phase space plots of a two-charge state beam along the LEBT: a) after $B_1$, b) before $B_2$, c) after $B_2$, d) RFQ entrance with scale changed to RFQ frequency.

## *2.2 RFQ*

The formation of heavy-ion beams of low longitudinal emittance has been discussed in ref. [8], in which the lowest beam emittance was obtained by prebunching and using a drift space inside the RFQ. This procedure works well for a single charge state beam, but cannot accommodate two charge states because of the different velocities of different charge states coming from the same ion source.

We describe below a new design for a low current RFQ and injector system which can provide very low longitudinal emittance for operation with both single charge state and two-charge state beams. Low rms and total longitudinal emittance are achieved by using an external multi-harmonic buncher and an RFQ of modified design. The RFQ has three main sections: 1) the standard radial matcher, 2) the transition section and 3) the acceleration section.

]The radial matcher transforms the RFQ acceptance to a set of Twiss parameters that avoids large beam sizes in the LEBT quadrupoles.

The transition section is a part of the RFQ with a linear variation of the synchronous phase. The RFQ parameters *A*, *m*, *a* and $\varphi_s$ are calculated self-consistently in order to filter the longitudinal emittance in such a way that the low populated area will be lost either inside the RFQ or in the MEBT. It is performed by an iterative procedure of the whole RFQ design and by observing the emittance formation in the longitudinal phase space. The desired result is achieved if the separatrix size is slightly larger than the densely populated area of beam emittance. The design goal is an rms emittance below 0.2 π·keV/u·nsec, required for multiple charge state operation through the rest of the linac.

The acceleration section is a portion of the RFQ with constant synchronous phase which is equal to –24°. The RFQ forms a beam crossover at the exit in both transverse planes. This results in better matching of the two-charge state beam in the MEBT.

In this way, the RFQ was designed for acceleration of one- or two-charge state heavy-ion beams from 12.4 keV/u to 167 keV/u over a length of 4 m. The phase space plots from numerical simulations of a two-charge state uranium beam exiting the RFQ are shown in Fig. 4.

## *2.3 Beam simulation in the MEBT and SRF linac*

Between the RFQ and SRF Linac there is a matching section – the MEBT. We found that SC solenoids placed in individual cryostats are the best option to focus a two-charge state beam.

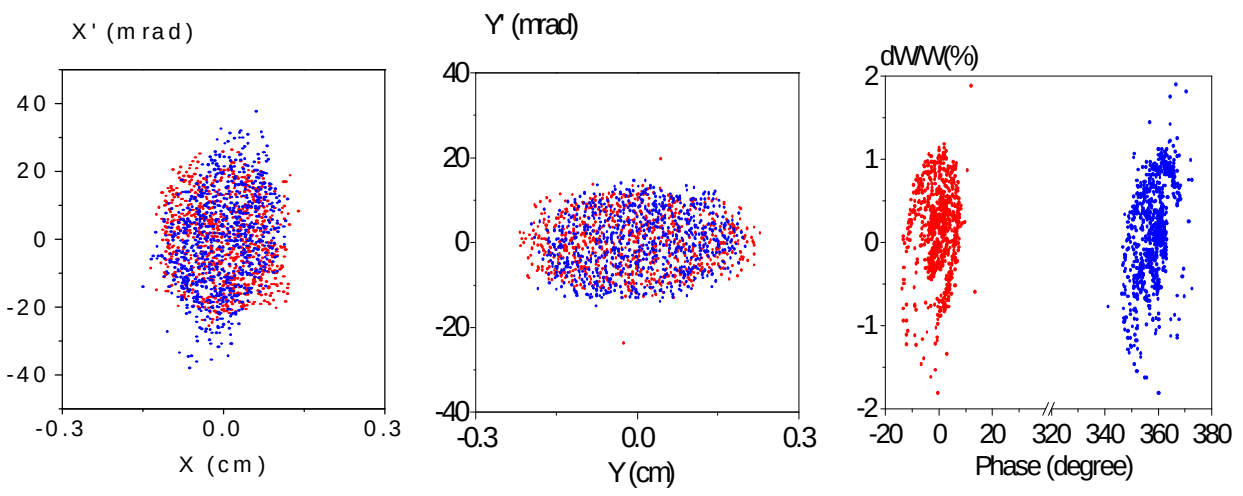

Figure 4: Phase space plots of beams with charge state 29+ and 30+ exiting the RFQ. In the transverse planes the charge states 29 and 30 occupy the same area, in the longitudinal plane the bunches are separated by 360°.

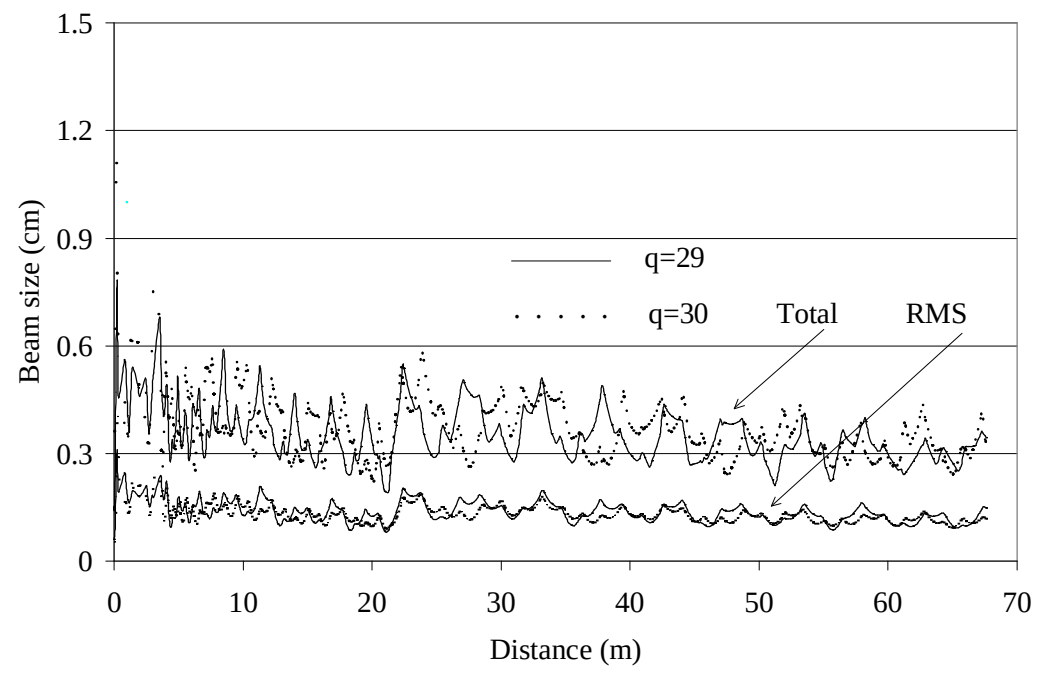

Figure 5: Two-charge state beam envelopes (rms and total) along the MEBT and SRF Linac.

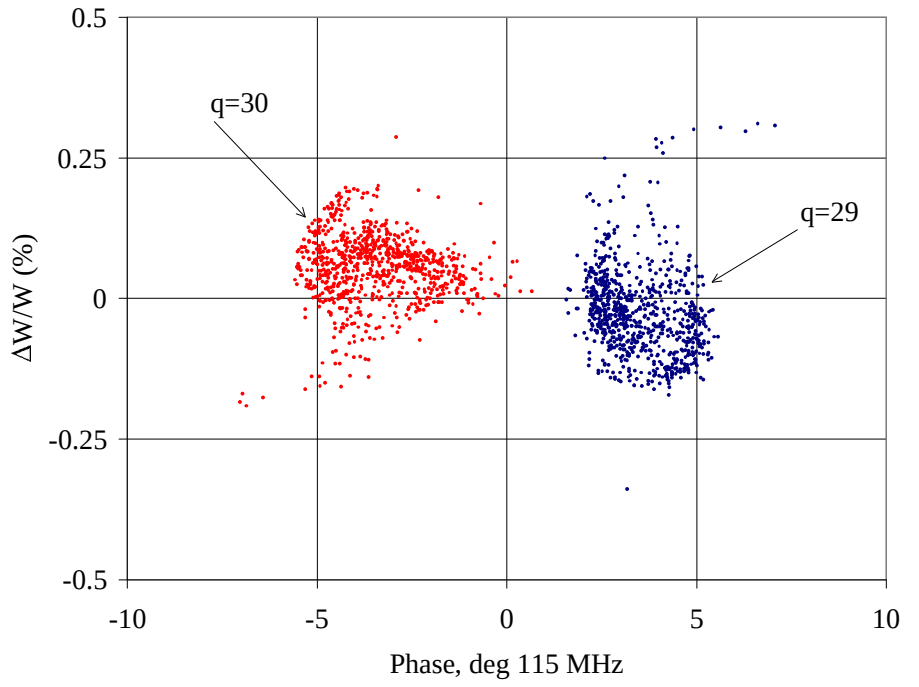

Figure 6: Phase space plots of a two-charge state beam just prior to the first stripper.

The portion of the linac prior to the first stripper contains 96 cavities of four different types. Prior to ray tracing a multiple charge state beam through the prestripper linac, the transverse beam motion was matched carefully using fitting codes for a trial beam of charge state q=29.5. A particularly critical aspect of fitting was to avoid beam mismatch at the transitions between focusing periods of differing lengths. The focusing lattice length is different for each of the four types of SRF cavities. The phase advance per focusing period was set at 60°. As mentioned in ref. [3], the correct choice of phase advance is crucial for effective steering of the multiple charge state beam. The design and simulation of 3D beam dynamics in the SRF linac was performed by the LANA code [9]. The pre-processor code generates the phase setting for a uranium beam with average charge state q=29.5. The rf phase is set to –30° with respect to the maximum energy gain in each SRF cavity. Realistic field distributions for the SRF cavities were generated using an axially-symmetric approximation of the actual drift tube cavities. These fields are used both for the design procedure and for the beam dynamics simulation. The initial phase space distribution used for each charge state was the beam at the exit of the RFQ, as simulated with the DYNAMION code. The particle coordinates were then tracked through the SRF linac. Figure 5 shows the transverse beam envelopes (rms and total) along the MEBT and SRF linac. Despite the slight mismatch of the two-charge state beam along the linac, there is no rms emittance growth in the transverse plane for an ideal linac without any errors.

In longitudinal phase space the emittance of the two charge state beam is always larger than for a single charge state beam. Growth in effective emittance occurs due to the oscillations caused by the slightly different, off-tune synchronous phases for charge states 29+ and 30+. However, the very low longitudinal emittance achieved by the RFQ injector ensures that the total emittance of the two-charge state beam remains well inside the stable area in longitudinal phase space. Figure 6 shows the longitudinal phase space just prior to the first stripper. Note that the energy and phase acceptance of the reminder of the SRF linac are ±3% and ±30°, respectively.

# 3 CONCLUSION

The problem of acceptance and acceleration of two charge states of a heavy-ion beam from a single ECR ion source was successfully solved. A front end has been designed for a driver linac for RIA that accelerates a two-charge state uranium beam. The use of a two-charge state beam is a powerful tool to double the total beam power produced by the heavy ion driver linac.

This work was supported by the U. S. Department of Energy under contract W-31-109-ENG-38